\documentclass[12pt]{article}

\usepackage{graphicx}
\usepackage{dcolumn}
\usepackage{bm}
\usepackage{amsmath}
\usepackage{amssymb}

\begin{document}

\begin{center}
\Large\bf\boldmath
\vspace*{0.8cm} SUSY constraints from relic density: high sensitivity \\
to pre-BBN expansion rate
\unboldmath
\end{center}

\vspace{0.6cm}
\begin{center}
A. Arbey$^{1,}$\footnote{Electronic address: \tt arbey@obs.univ-lyon1.fr} and F. Mahmoudi$^{2,}$\footnote{Electronic address: \tt nazila.mahmoudi@tsl.uu.se}\\[0.4cm]
{\sl $^1$ Universit\'e de Lyon, Lyon, F-69000, France; Universit\'e Lyon~1, Villeurbanne, \\
F-69622, France; Centre de Recherche Astrophysique de Lyon, Observatoire de Lyon, \\
9 avenue Charles Andr\'e, Saint-Genis Laval cedex, F-69561, France; CNRS, UMR 5574; \\
Ecole Normale Sup\'erieure de Lyon, Lyon, France.}\\
\vspace{0.6cm}
{\sl $^2$ High Energy Physics, Uppsala University, Box 535, 75121 Uppsala, Sweden.}
\end{center}

\vspace{0.4cm}
\begin{abstract}
\noindent 
The sensitivity of the lightest supersymmetric particle relic density calculation to the variation of the cosmological expansion rate before nucleosynthesis is discussed. We show that such a modification, even extremely modest and with no consequence on the cosmological observations, can greatly enhance the calculated relic density, and therefore change the constraints on the SUSY parameter space drastically. We illustrate this variation in two examples of SUSY models, and show that it is unsafe to use the lower bound of the WMAP limits in order to constrain supersymmetry. We therefore suggest to use only the upper value $\Omega_{DM} h^2 < 0.135$.
\end{abstract}
%

\vspace{1.2cm}
\noindent During the past decade, supersymmetry (SUSY), as one of the most promising candidates for new physics beyond the Standard Model, has been the focus of intensive phenomenological studies. Huge efforts have been carried out in order to constrain the supersymmetric parameter space. Among the most powerful observables for this purpose, in addition to the direct searches at LEP and Tevatron, stand the WMAP limits on the relic density and $B$ physics constraints.\\
In this letter, we present a new analysis of the relic density constraints on the Minimal Supersymmetric extension of the Standard Model (MSSM), and we focus in particular on bounds on two gravity mediated supersymmetry breaking scenarios, namely minimal Supergravity (mSUGRA) and the Non-Universal Higgs Mass framework (NUHM) in which the boundary conditions at high scales reduce the number of free parameters of the MSSM, allowing feasible phenomenological studies.\\
\\
The recent observations of the WMAP satellite \cite{wmap}, combined with other cosmological data, give evidence for the presence of a cosmological matter-like density representing about 27\% of the total density of the Universe. The remaining 73\% reveal the presence of the so-called dark energy. From the total matter density observed by WMAP \cite{wmap} and the baryon density indicated by Big-Bang nucleosynthesis (BBN) \cite{nucleo}, including theoretical uncertainties, the dark matter density range at 95\% C.L. can be deduced:
\begin{equation}
 0.094 < \Omega_{DM} h^2 < 0.135 \;, \label{WMAPnew}
\end{equation}
where $h$ is the reduced Hubble constant. In the following, we also refer to the older range \cite{ellis97} which admits a larger interval:
\begin{equation}
 0.1 < \Omega_{DM} h^2 < 0.3 \;. \label{WMAPold}
\end{equation}
The lightest supersymmetric particle (LSP), provided it is stable and electrically neutral, constitutes the favorite candidate for non-baryonic dark matter. The stability requirement is fulfilled when R-parity is conserved, and scenarios such as mSUGRA or NUHM provide us with a LSP satisfying the WMAP relic density constraints \cite{ellis03}.\\
The great accuracy of the WMAP data can therefore be used to constrain the supersymmetric parameters, provided the relic density is calculated precisely. The computation of the relic density has been realized within the standard model of cosmology \cite{edsjo97}, and implemented in automatic codes, such as MicrOMEGAs \cite{micromegas} and DarkSUSY \cite{darksusy}.\\
However, in the standard model of cosmology, the nature of the dark energy and the evolution of the Universe in the pre-BBN era remain unclear. The BBN era is the oldest period in the cosmological evolution when reliable constraints are derived, for temperatures of about 1 MeV. Successful BBN models predict that radiation was the dominant energy at that time, but no claim is made for much higher temperatures. In fact, in models like quintessence \cite{quintessence}, $k$-essence \cite{Kessence} or dark fluid \cite{DF}, dark energy could play a role before BBN, since its density could be much higher at such temperatures, as was especially underlined within the quintessence model in \cite{salati03}. Also, some extra-dimension theories predict negative effective energies in the Early Universe, which can modify the relic density \cite{negative_density}.\\
Therefore, the standard model of cosmology could be more complex than what we think in the primordial Universe, and the pre-BBN era could have experienced a slower or faster expansion. Such a modified expansion, even though still compatible with the BBN or the WMAP results, changes the LSP freeze-out time and the amount of relic density.\\
\\
To model the effects of such a modified expansion in the pre-BBN era, we add to the radiation density a new dark density, varying with temperature as
\begin{equation}
\rho_D(T) = \rho_D(T_0) \left(\frac{T}{T_0}\right)^{n_D}\;,
\end{equation}
where we choose $T_0=1$ MeV and $n_D$ is a constant parametrizing the density behavior. Such a density evolution characterizes a fluid in adiabatic expansion with a constant equation of state $w_D=P_D/\rho_D$, where $P_D$ is the pressure of the fluid: for $n_D=3$ ($w_D=0$) the dark density evolves as a matter density; for $n_D=4$ ($w_D=1/3$) as a radiation density; and for $n_D=6$ ($w_D=1$) as the density of a real scalar field ({\it e.g.} a quintessence field) with a dominating kinetic term \cite{quintessence}. $n_D>6$ can arise for example in extra-dimension models.
We introduce the parameter
\begin{equation}
\kappa_D \equiv \frac{\rho_D(T_0)}{\rho_{\mbox{rad}}(T_0)}\;,
\end{equation}
where $\rho_{\mbox{rad}}$ is the radiation density, evolving as
\begin{equation}
\rho_{\mbox{rad}} (T) = g_{\mbox{eff}}(T)\frac{\pi^2}{30} T^4\;.
\end{equation}
$g_{\mbox{eff}}$ is the effective number of degrees of freedom of the radiation. $\kappa_D$ parametrizes the temperature at which the dark density dominates the expansion, {\it i.e.} $\rho_D(T) > \rho_{\mbox{rad}}(T)$; the larger $\kappa_D$ is, the earlier the dark density dominates. In particular, if $\kappa_D=1$, the radiation and the dark component will be co-dominant at BBN time. Thus, imposing the radiation density to remain dominant at BBN time and later leads to
\begin{equation}
n_D \ge 4 \;\;\;\mbox{and}\;\;\; |\kappa_D| < 1 \;.
\end{equation}
For a usual scalar field $0 \le n_D \le 6$, but with a modified kinetic term $n_D$ can reach higher values. We restrict here however to $n_D \lesssim 8$. Furthermore, if $\kappa_D < 0$, the extra requirement $\rho_D+\rho_{\mbox{rad}}>0$ should be satisfied at any time, which limits negative effective densities to $n_D \approx 4$ or to very low $|\kappa_D|$.\\
The Friedmann equation at BBN time and before reads
\begin{equation}
H^2=\frac{8\pi G}{3} (\rho_{\mbox{rad}} + \rho_D) \;, \label{friedmann}
\end{equation}
and the dynamics of the expansion is therefore modified, leading to a higher expansion rate if $\rho_D>0$, or a lower one if $\rho_D<0$. Here $\rho_D$ does not necessarily have to correspond to the density of a real component, but can be only an effective term to parametrize the modification of the expansion rate.\\
Under the standard hypotheses, {\it i.e.} in absence of entropy production and of nonthermal generation of relic particles, the computation of the relic density is based on the solution of the evolution equation \cite{edsjo97}
\begin{equation}
\frac{dn}{dt}=-3Hn-\langle \sigma_{\mbox{eff}} v\rangle (n^2 - n_{\mbox{eq}}^2)\;, \label{evol_eq}
\end{equation}
where $n$ is the number density of all supersymmetric particles, $n_{\mbox{eq}}$ their equilibrium density, and $\langle \sigma_{\mbox{eff}} v\rangle$ is the thermal average of the annihilation rate of the supersymmetric particles to Standard Model particles. By solving this equation, the density number of supersymmetric particles in the present Universe and consequently the relic density can be determined.\\
We consider the ratio of the number density to the radiation entropy density,
\begin{equation}
Y(T)=\frac{n(T)}{s(T)} \;,
\end{equation}
with
\begin{equation}
s(T)=h_{\mbox{eff}}(T) \frac{2 \pi^2}{45} T^3 \;.
\end{equation}
$h_{\mbox{eff}}$ is the effective number of entropic degrees of freedom of radiation. Combining Eqs. (\ref{friedmann}) and (\ref{evol_eq}) and defining $x=m_{\mbox{\small LSP}}/T$, the ratio of the LSP mass over temperature, yield
\begin{equation}
\frac{dY}{dx}=-\sqrt{\frac{\pi}{45 G}}\frac{g_*^{1/2} m_{\mbox{\small LSP}}}{x^2} \left(1+\frac{\rho_D(T)}{g_{\mbox{eff}}(T) \frac{\pi^2}{30}T^4 } \right)^{-1/2} \langle \sigma_{\mbox{eff}} v\rangle (Y^2 - Y^2_{\mbox{eq}}) \;, \label{main}
\end{equation}
with
\begin{equation}
g_*^{1/2}=\frac{h_{\mbox{eff}}}{\sqrt{g_{\mbox{eff}}}}\left(1+\frac{T}{3 h_{\mbox{eff}}}\frac{dh_{\mbox{eff}}}{dT}\right) \;.
\end{equation}
Note that in the limit where $\rho_D \to 0$, we retrieve the results of Ref. \cite{edsjo97}.\\
\\
The freeze-out temperature $T_f$ is the temperature at which the LSP leaves the initial thermal equilibrium, {\it i.e.} $T=T_f$ when $Y (T_f) = (1 + \delta) Y_{\mbox{eq}}(T_f)$, with $\delta \simeq 1.5$.
The relic density is obtained by integrating Eq. (\ref{main}) from $x=0$ to $m_{\mbox{\small LSP}}/T_0$, where $T_0=2.726$ K is the temperature of the Universe today \cite{edsjo97}:
\begin{equation}
\Omega_{\mbox{\small LSP}} h^2 = 2.755\times 10^8 \frac{m_{\mbox{\small LSP}}}{1 \mbox{ GeV}} Y(T_0)\;.
\end{equation}
To compute numerically the relic density, we use a modified version of MicrOMEGAs 2.0.7 \cite{micromegas} which includes the alteration of the expansion rate in the primordial Universe, as in Eq. (\ref{friedmann}). The SUSY mass spectrum and couplings are computed with SOFTSUSY 2.0.14 \cite{softsusy}, and the $b\to s \gamma$ branching ratio and isospin asymmetry are calculated with SuperIso 2.0 \cite{superiso}, using the limits of \cite{mahmoudi07}.\\
\\
\begin{figure}[!t]
\begin{center}
\includegraphics[height=7cm]{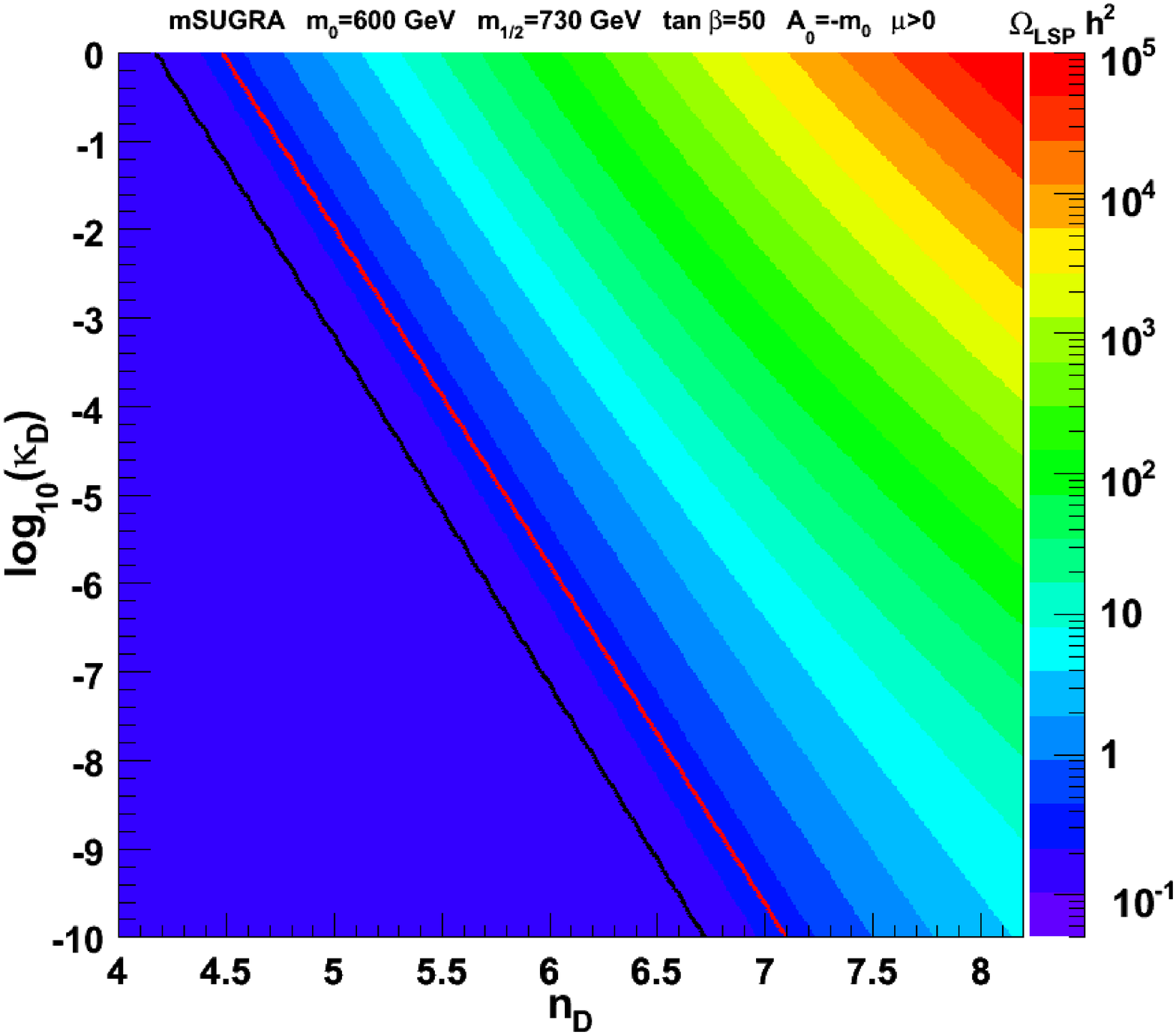}\includegraphics[height=7cm]{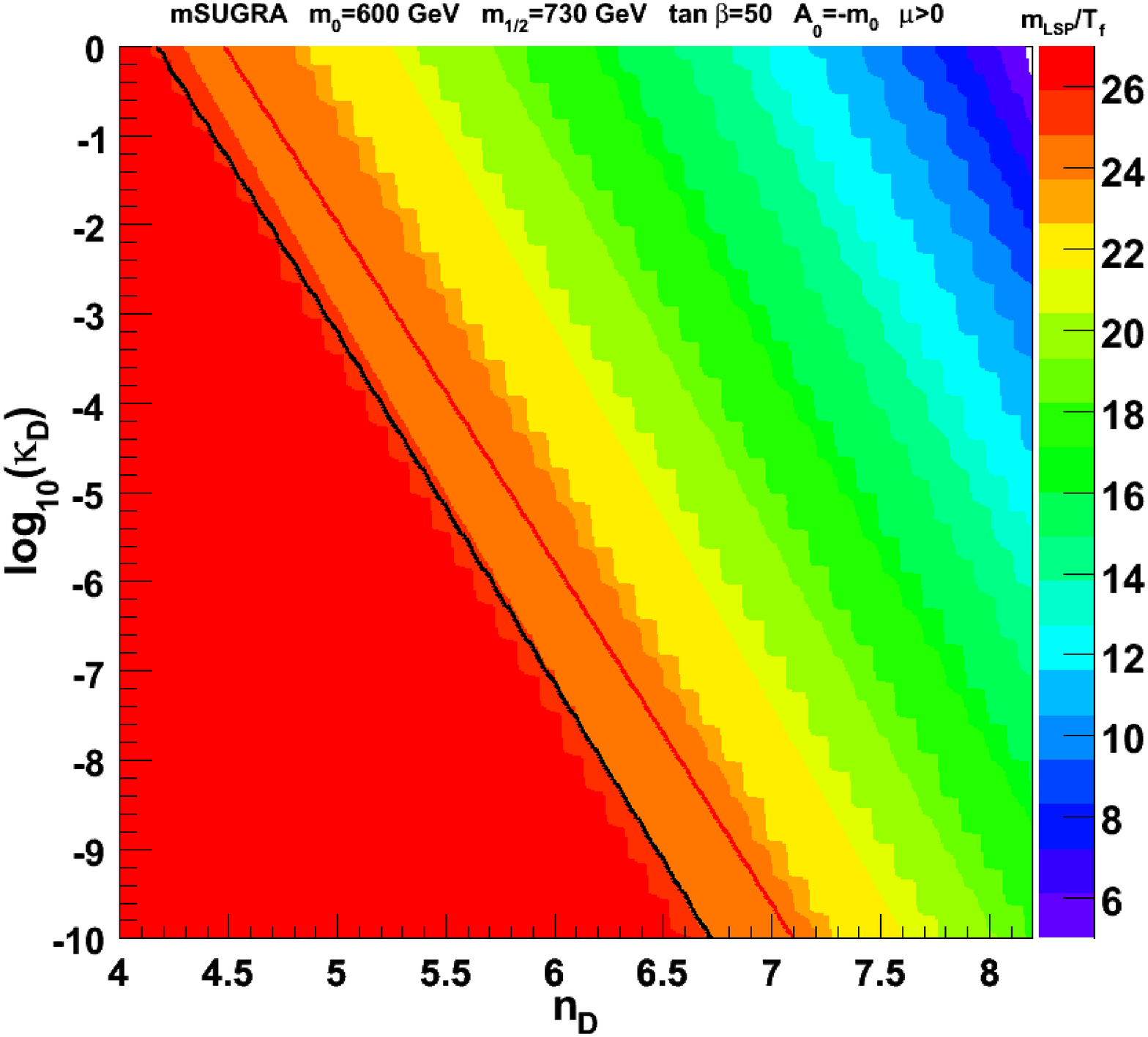}
\end{center}
\caption{Relic density (left) and freeze-out temperature (right) in the modified pre-BBN expansion model for the mSUGRA point specified in the text. The regions on the right of the oblique black/red lines are disfavored by the new/old constraints of (\ref{WMAPnew}) and (\ref{WMAPold}).\vspace*{1.0cm}}\label{fig1}
\end{figure}%
Let us consider first the mSUGRA parameter point ($m_0=600$ GeV, $m_{1/2}=730$ GeV, $A_0=-m_0$, $\tan\beta=50$, $\mu>0$), which is favored by WMAP, as an example. The standard calculation of the relic density for this point leads to  $m_{\mbox{\small LSP}} = 308$ GeV, $\Omega_{\mbox{\small LSP}} h^2 = 0.105$, and a freeze-out temperature of $T_f = 11$ GeV. In Fig. \ref{fig1}, the dependence of the relic density and of the freeze-out temperature on $n_D$ and $\kappa_D$ is shown. The oblique black and red lines correspond, respectively, to the present upper limit of the WMAP constraint $\Omega_{\mbox{\small LSP}} h^2 = 0.135$, and to the older limit $\Omega_{\mbox{\small LSP}} h^2 = 0.3$. $\kappa_D$ is varied in the interval $[10^{-10},1]$, and $n_D$ in $[4,8.2]$.\\
First we note that the relic density can be increased by up to a factor $10^6$ and the freeze-out temperature up to 50 GeV. This strong dependence of the relic density and freeze-out temperature on $n_D$ and $\kappa_D$ is due to the fact that for large values of $n_D$ and $\kappa_D$, the Universe is not anymore dominated by the radiation at freeze-out temperature, but by the dark fluid. Avoiding the extreme values of $n_D$ and $\kappa_D$, and considering for example a quintessence fluid with $n_D = 6$, we observe that even with a negligible contribution of dark fluid at BBN time, e.g. $\kappa_D\sim 10^{-3}$, the relic density is multiplied by a factor larger than 100, and this point is then excluded by the WMAP limits. For lower values of $n_D$ and of $\kappa_D$, the relic density remains however inside the favored interval.\\
For negative effective densities the calculated relic density hardly changes. With $n_D=4$, it decreases by less than 1\%, even with the excessive value $\kappa_D=-0.5$.\\
\\
We study now the effects of a quintessence density with $n_D=6$ and $\kappa_D \in [10^{-5}, 10^{-2}]$, and of a density with $n_D=8$ and $\kappa_D=10^{-5}$, in the mSUGRA and NUHM parameter spaces. Note that this choice of parameters is completely in agreement with the known cosmological constraints, and such a modification of expansion rate would be transparent on usual observables. Nevertheless, it would affect the constraints obtained on the SUSY parameter spaces.\\
\\
In Fig. \ref{fig2}, we present the constraints on the mSUGRA parameter plane $(m_{1/2}, m_0)$, for $\tan\beta=50$, $A_0=-m_0$ and $\mu>0$. In the yellow region the LSP is charged, therefore this region is cosmologically disfavored. The green region is disfavored by the branching ratio of $B \to X_s \gamma$, the red region by the isospin asymmetry of $B \to K^* \gamma$, the gray area is excluded by the collider constraints on the particles masses \cite{PDG}, and the dark blue and light blue are respectively favored by the new and old constraints of (\ref{WMAPnew}) and (\ref{WMAPold}). All contours are at 95\% C.L. The top-left plot corresponds to the relic density calculated with the usual expansion rate. From top to bottom and left to right $\kappa_D$ is increased (as well as $n_D$ for the last plot), and we can notice the change in the relic density favored contours.\\
\begin{figure}[!p]
\begin{center}
\includegraphics[height=6cm]{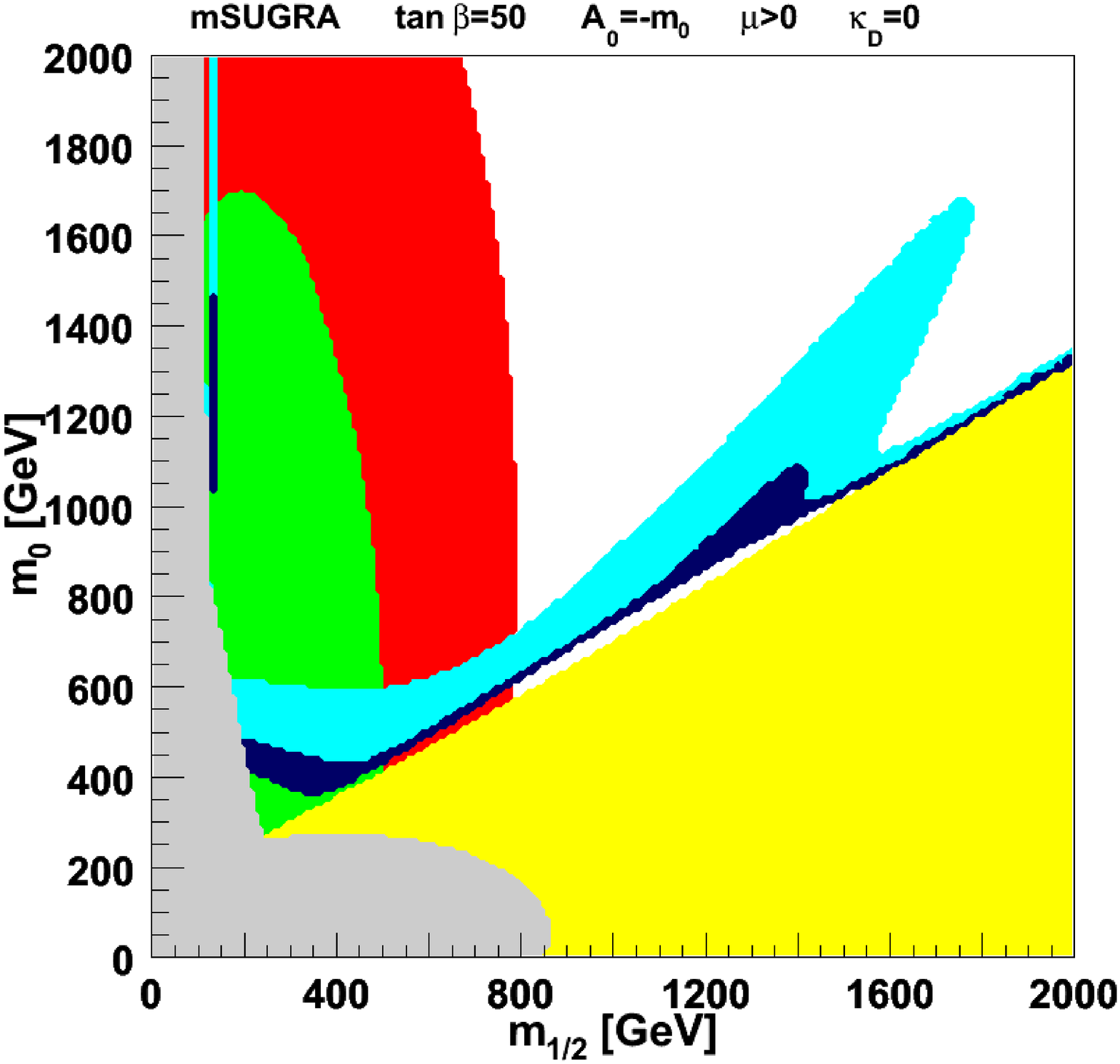}\includegraphics[height=6cm]{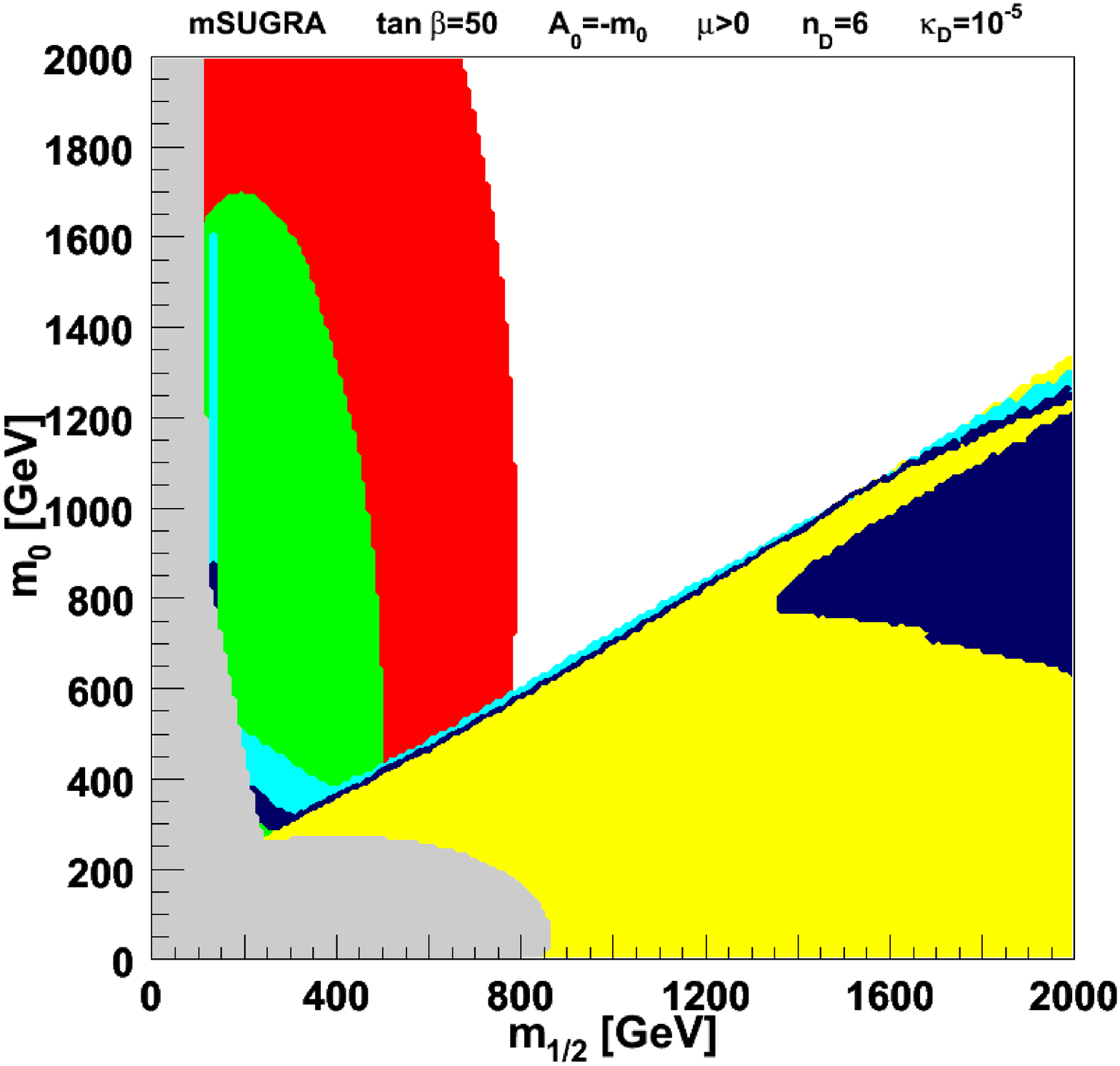}\\
\includegraphics[height=6cm]{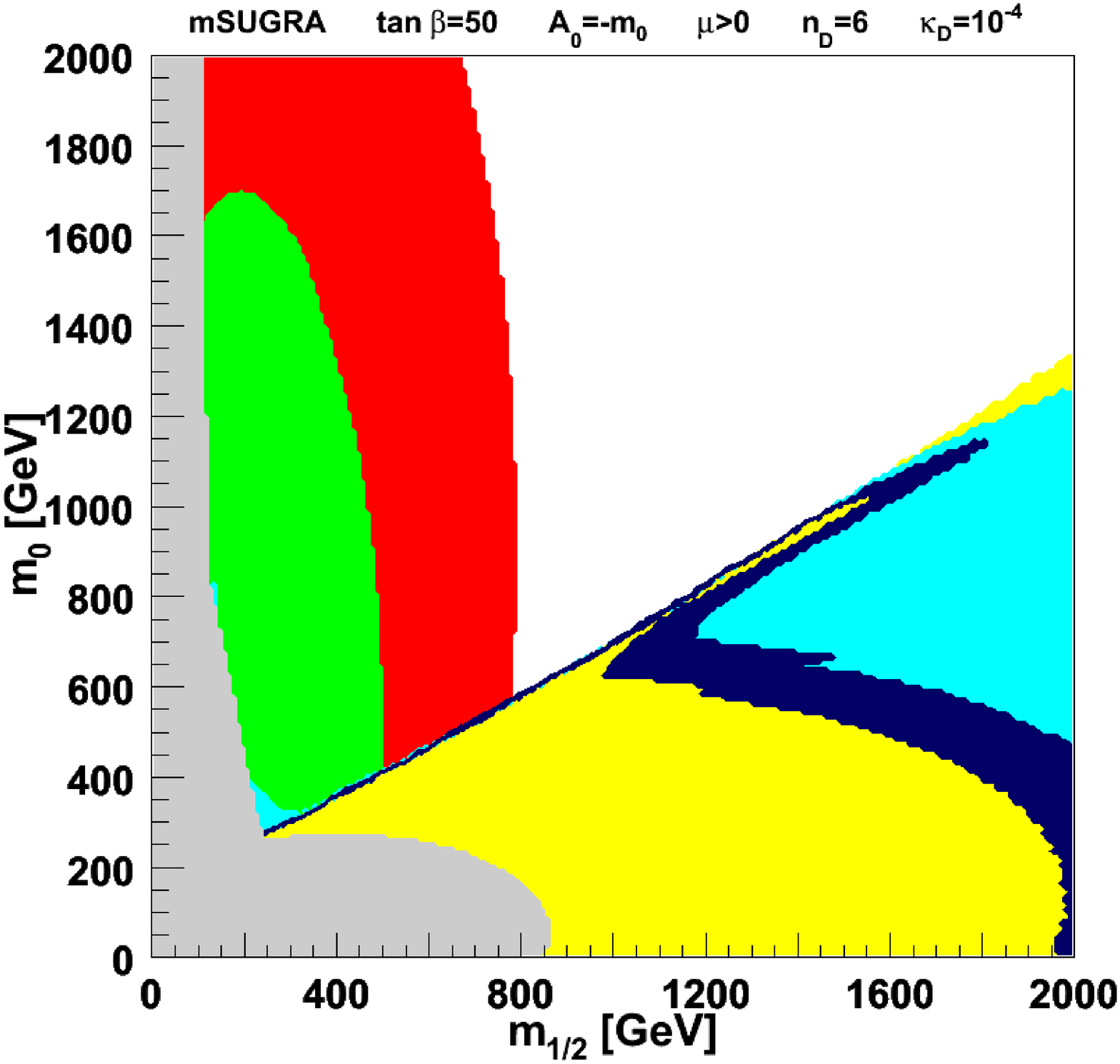}\includegraphics[height=6cm]{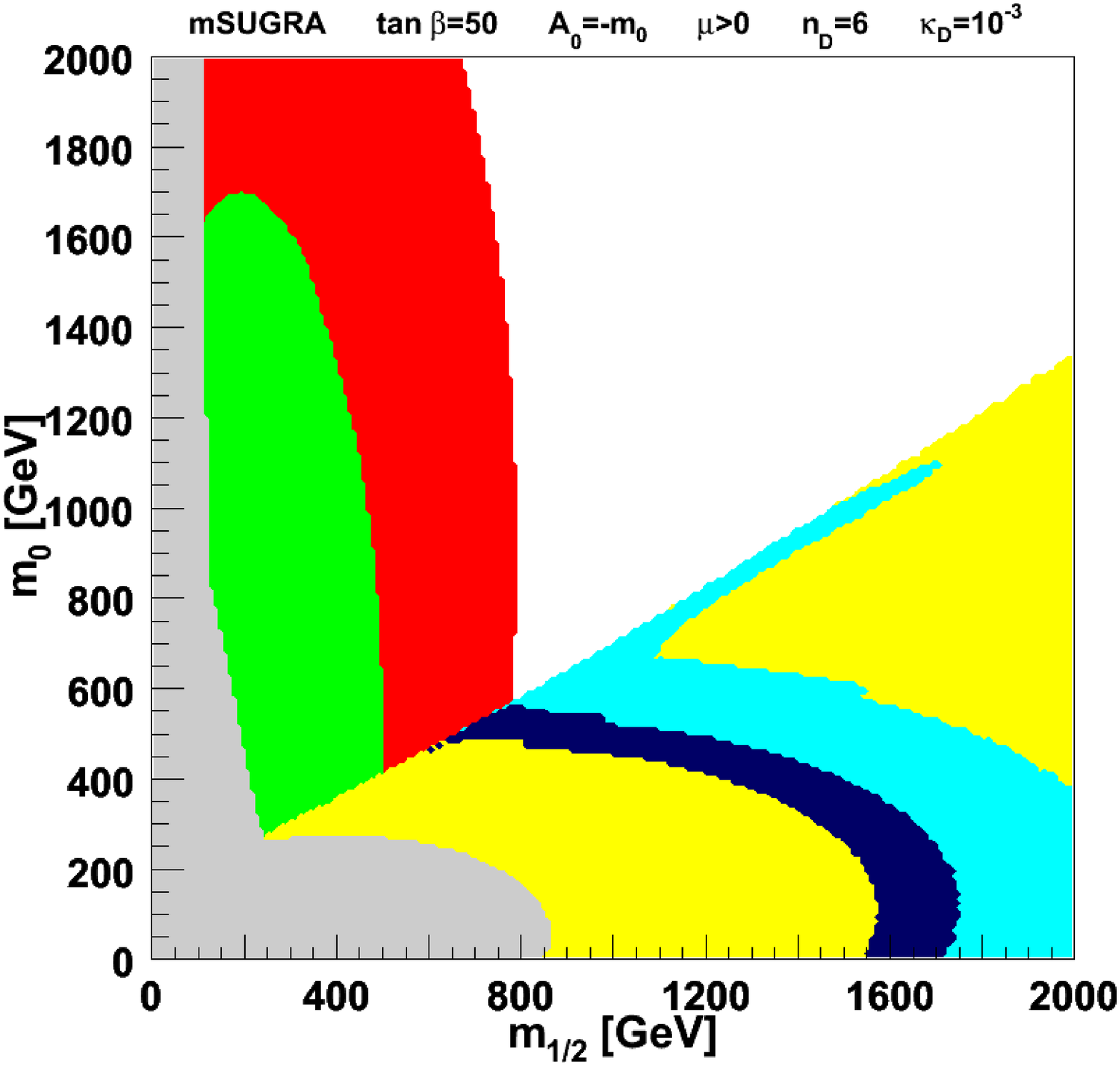}\\
\includegraphics[height=6cm]{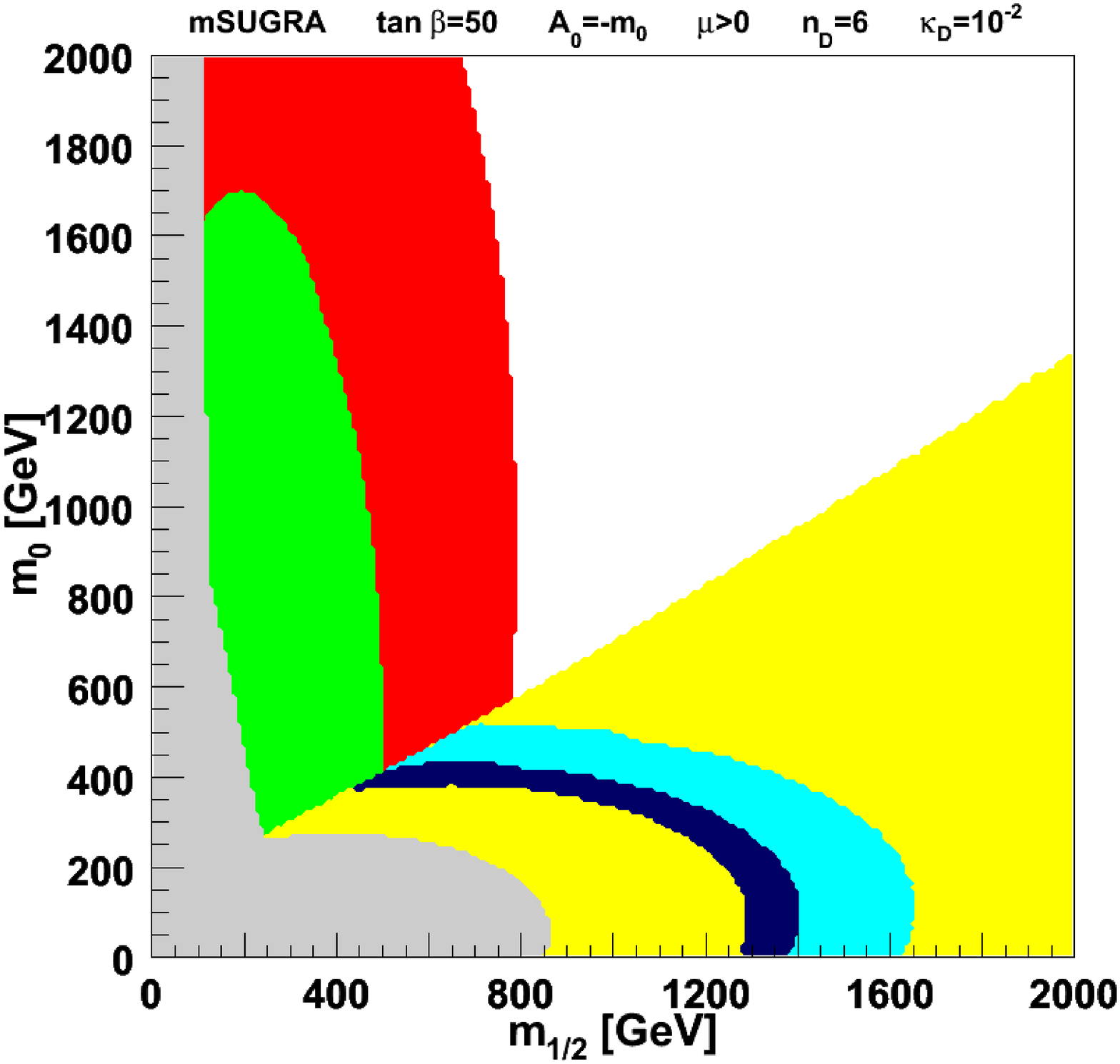}\includegraphics[height=6cm]{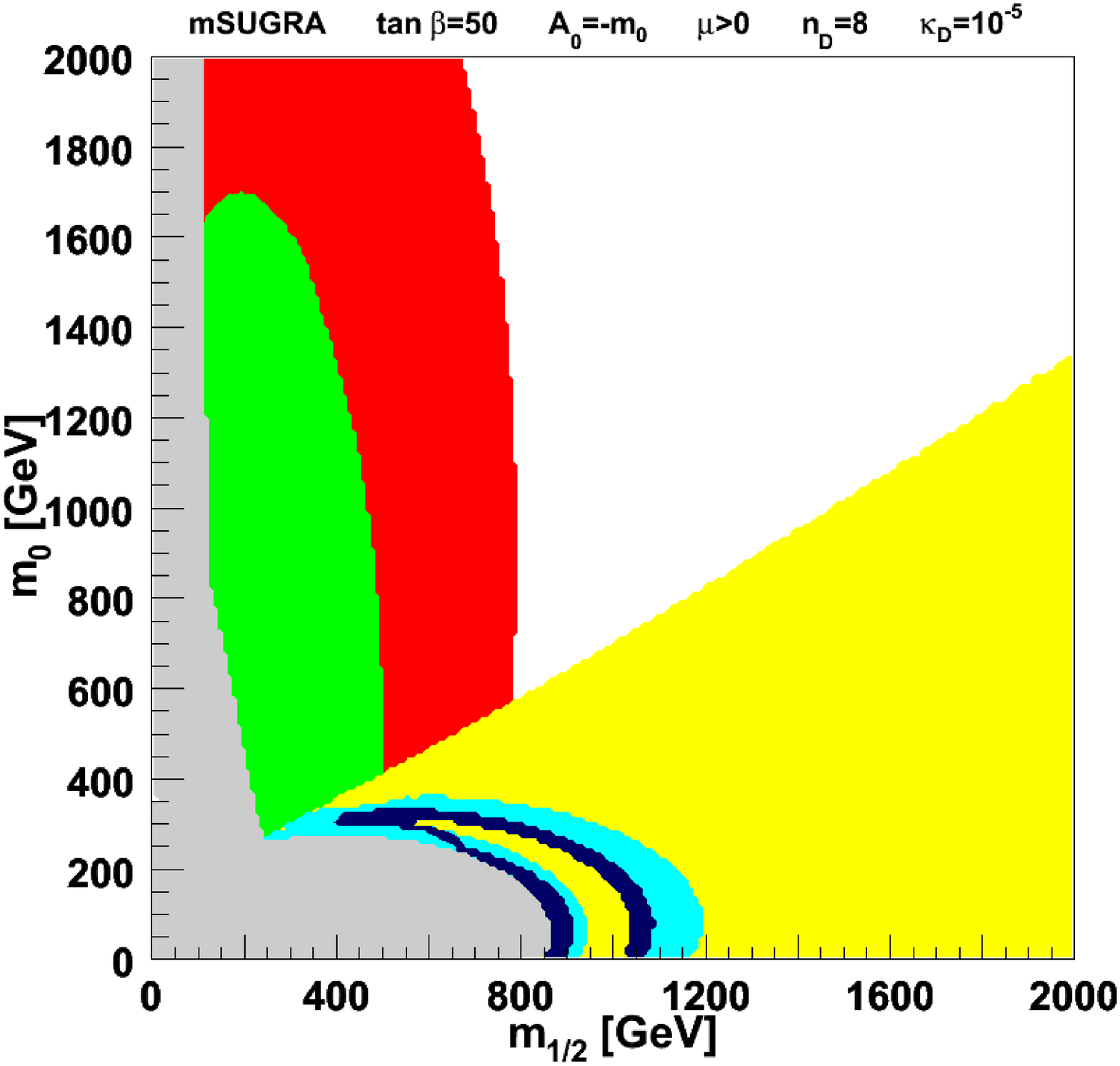}
\end{center}
\vspace*{-0.5cm}\caption{Constraints in the mSUGRA parameter plane for several values of $\kappa_D$ and $n_D$: $n_D=0$, $\kappa_D=0$ (top left), $n_D=6$, $\kappa_D=10^{-5}$ (top right), $n_D=6$, $\kappa_D=10^{-4}$ (middle left), $n_D=6$, $\kappa_D=10^{-3}$ (middle right), $n_D=6$, $\kappa_D=10^{-2}$ (bottom left), $n_D=8$, $\kappa_D=10^{-5}$ (bottom right). The dark and light blues correspond respectively to the regions favored by the new and the old dark matter constraints. The other colors refer to exclusion contours as described in the text.}\label{fig2}
\end{figure}%
The first important feature of this figure is that even with a modest modification of the expansion rate, a large part of the favored zone enters into the charged LSP area. Only a very narrow line is still favored for $\kappa_D=10^{-4}-10^{-5}$ and $n_D=6$, but becomes completely excluded for larger $\kappa_D$ and $n_D$. Therefore, a slight alteration of the pre-BBN expansion rate can induce drastic modifications in the mSUGRA WMAP favored regions.\\
\\
\begin{figure}[!p]
\begin{center}
\includegraphics[height=6cm]{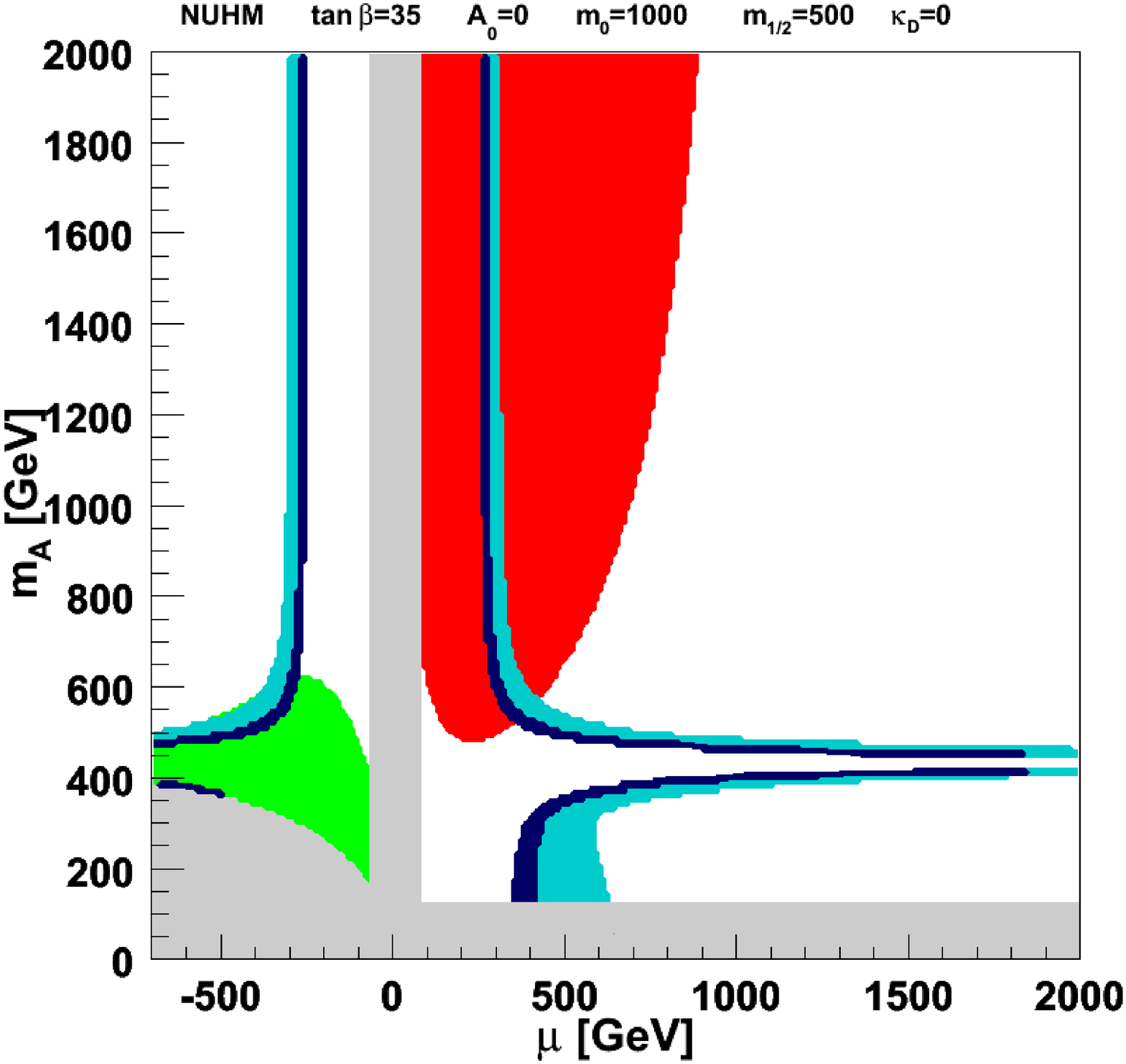}\includegraphics[height=6cm]{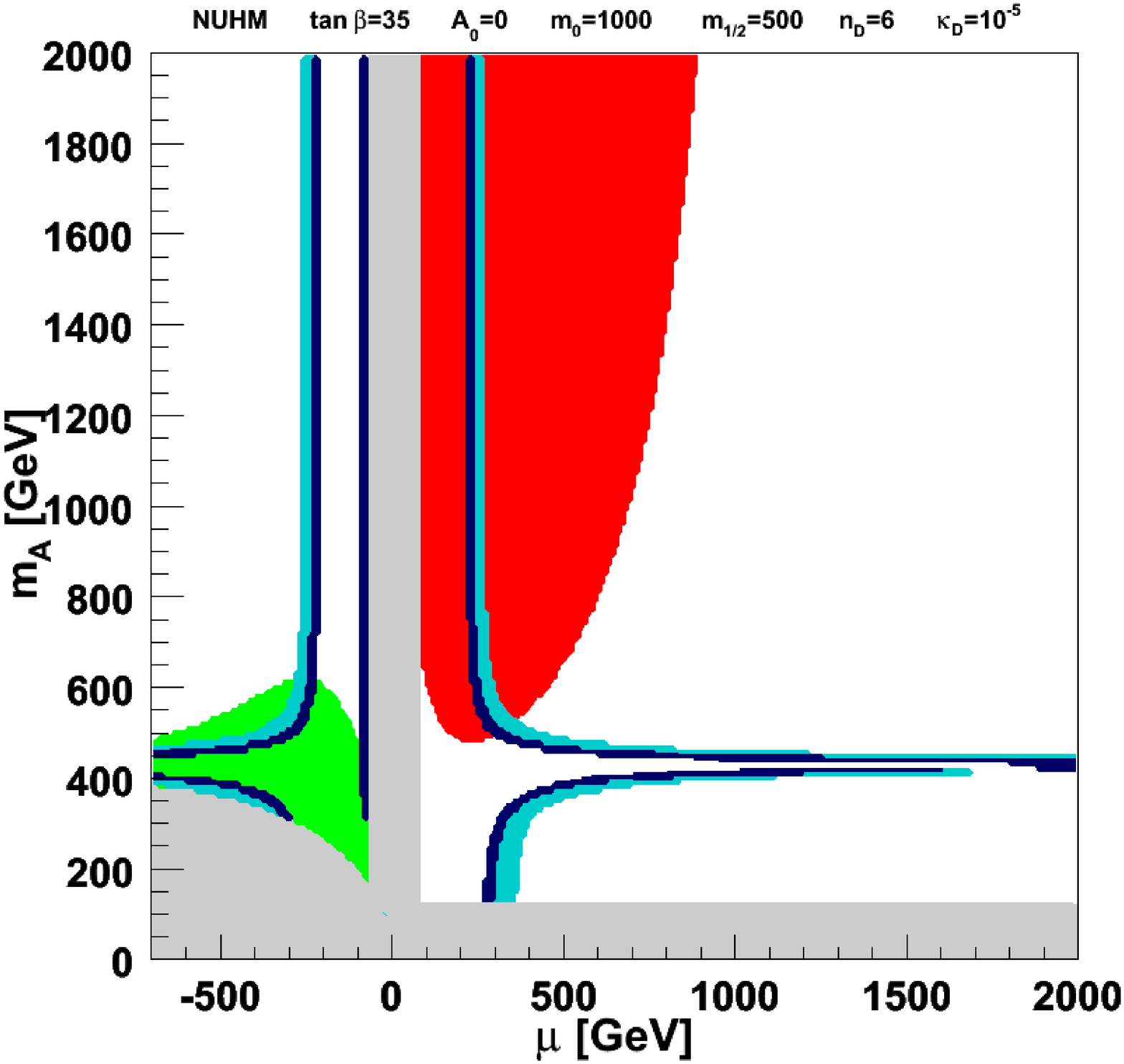}\\
\includegraphics[height=6cm]{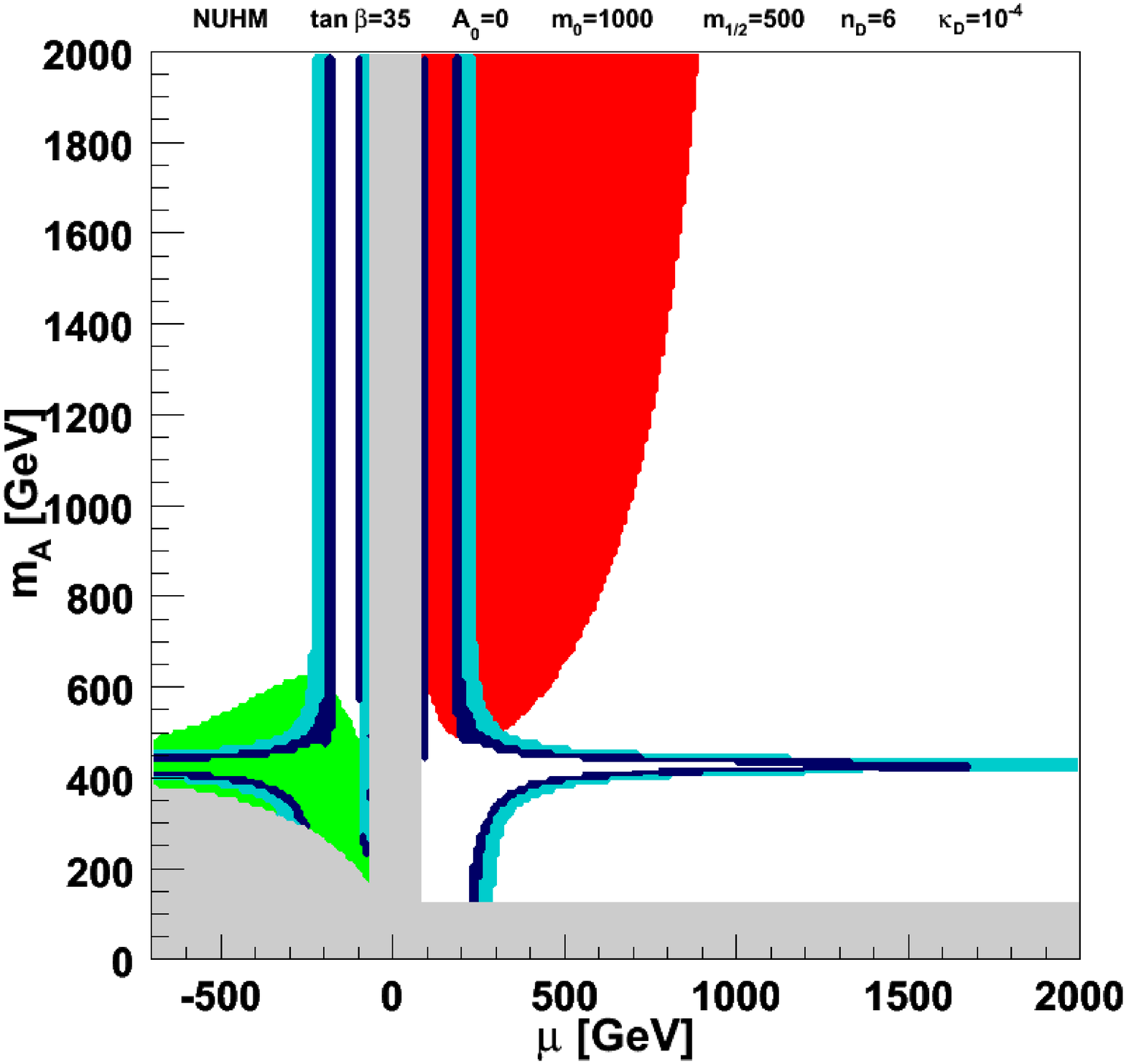}\includegraphics[height=6cm]{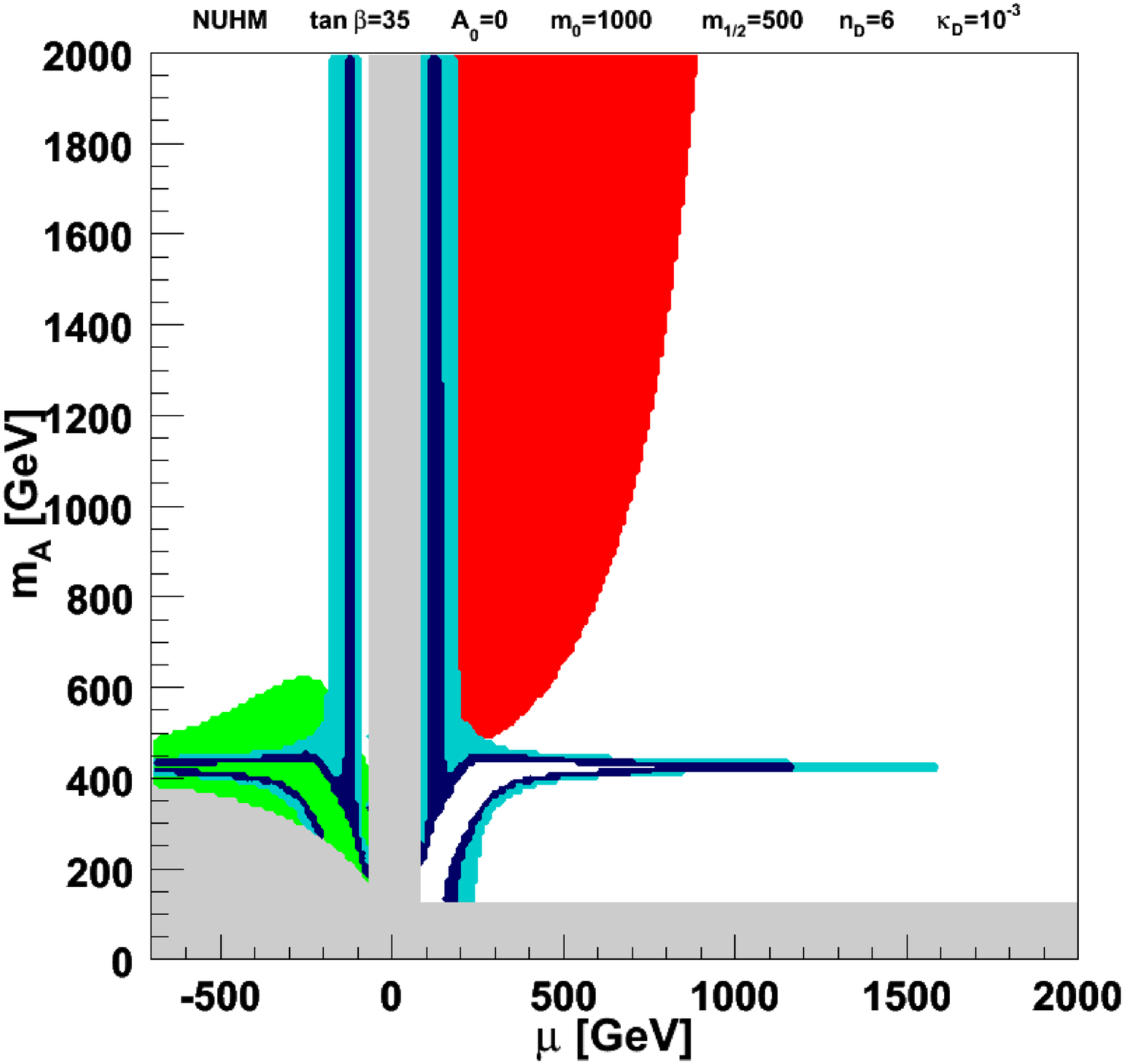}\\
\includegraphics[height=6cm]{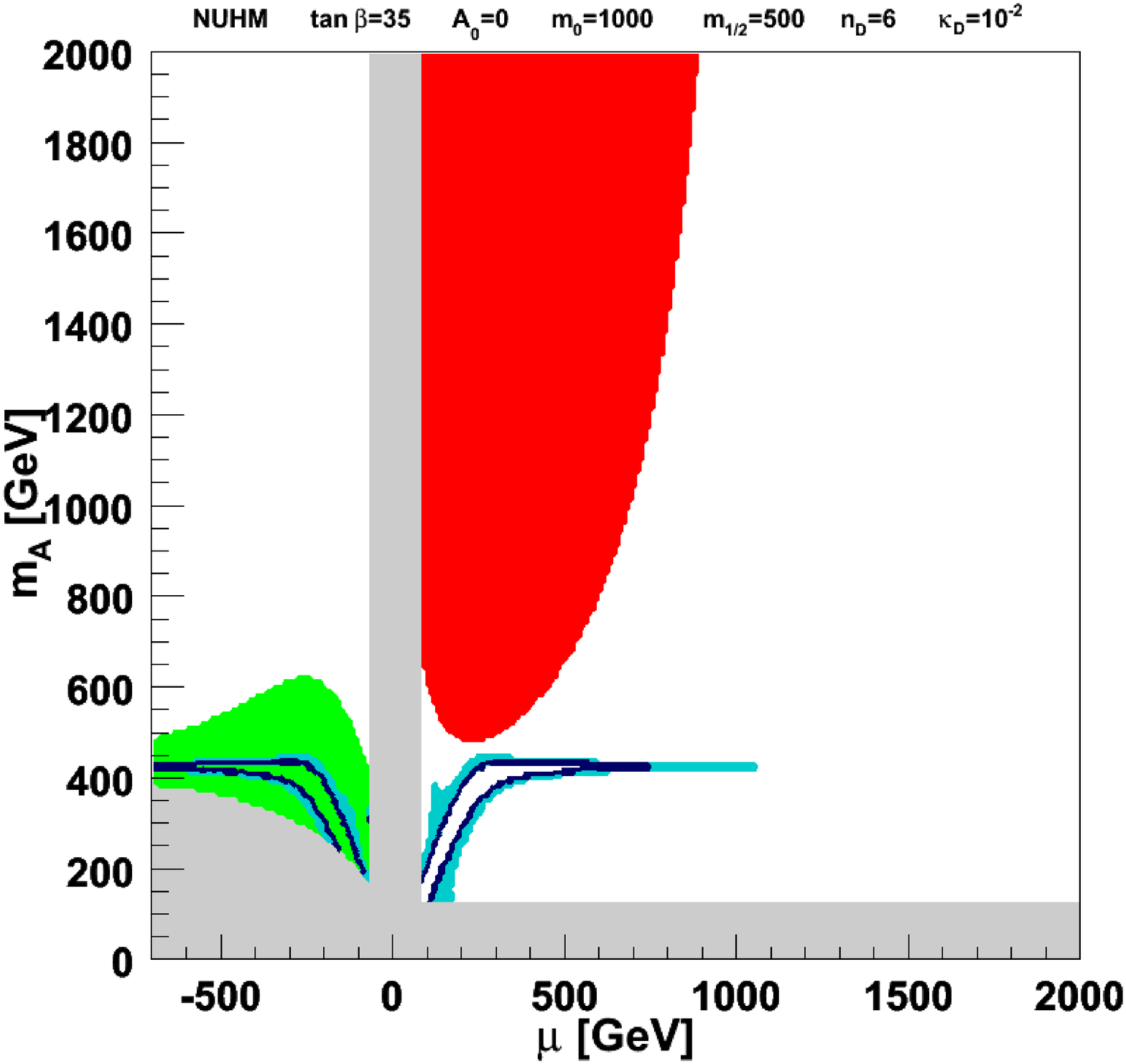}\includegraphics[height=6cm]{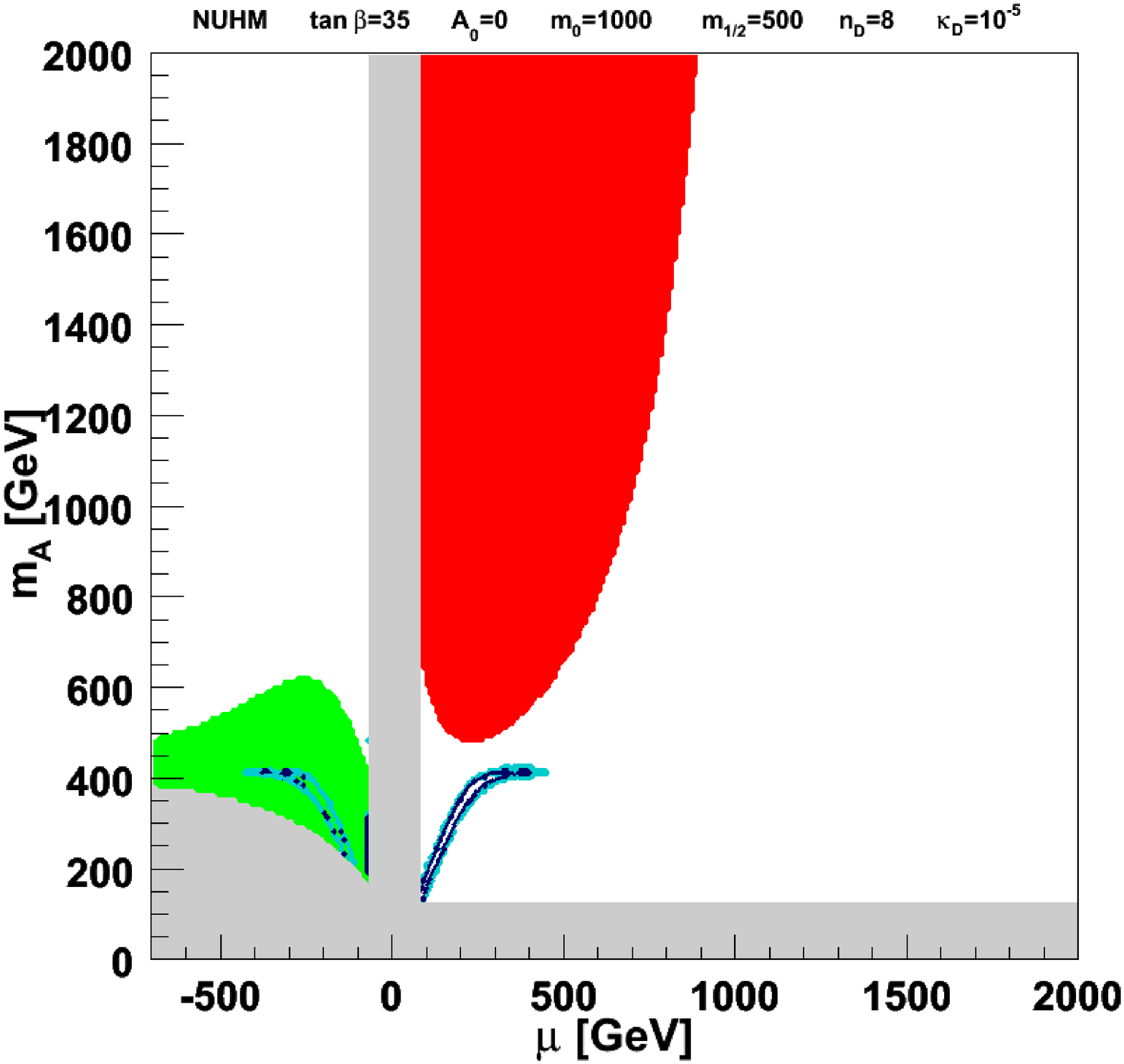}
\end{center}
\caption{Constraints in the NUHM parameter plane for several values of $\kappa_D$ and $n_D$: $n_D=0$, $\kappa_D=0$ (top left), $n_D=6$, $\kappa_D=10^{-5}$ (top right), $n_D=6$, $\kappa_D=10^{-4}$ (middle left), $n_D=6$, $\kappa_D=10^{-3}$ (middle right), $n_D=6$, $\kappa_D=10^{-2}$ (bottom left), $n_D=8$, $\kappa_D=10^{-5}$ (bottom right). The color conventions are described in the text.}\label{fig3}
\end{figure}%
\begin{figure}[!t]
\begin{center}
\includegraphics[height=7cm]{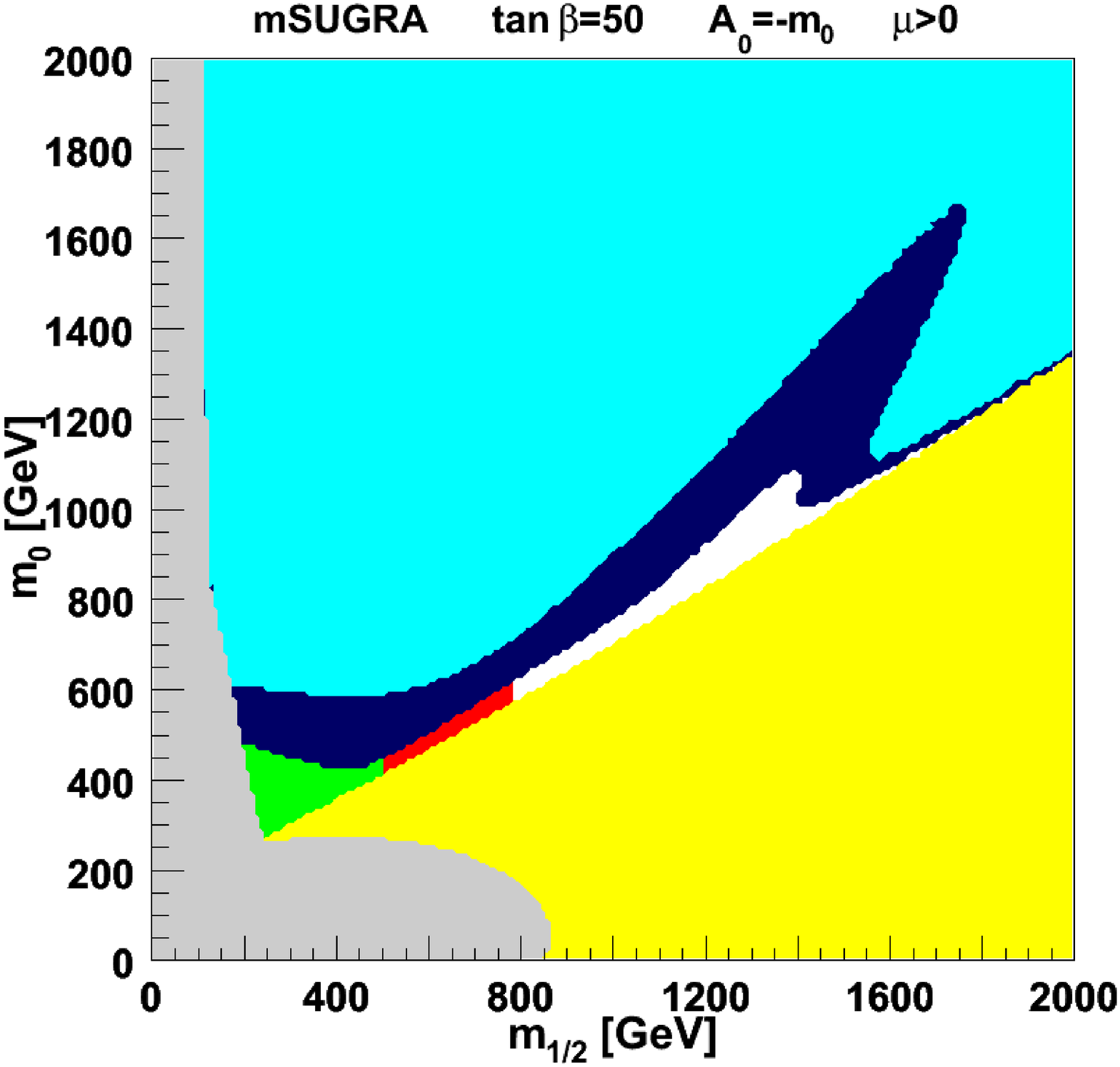}\includegraphics[height=7cm]{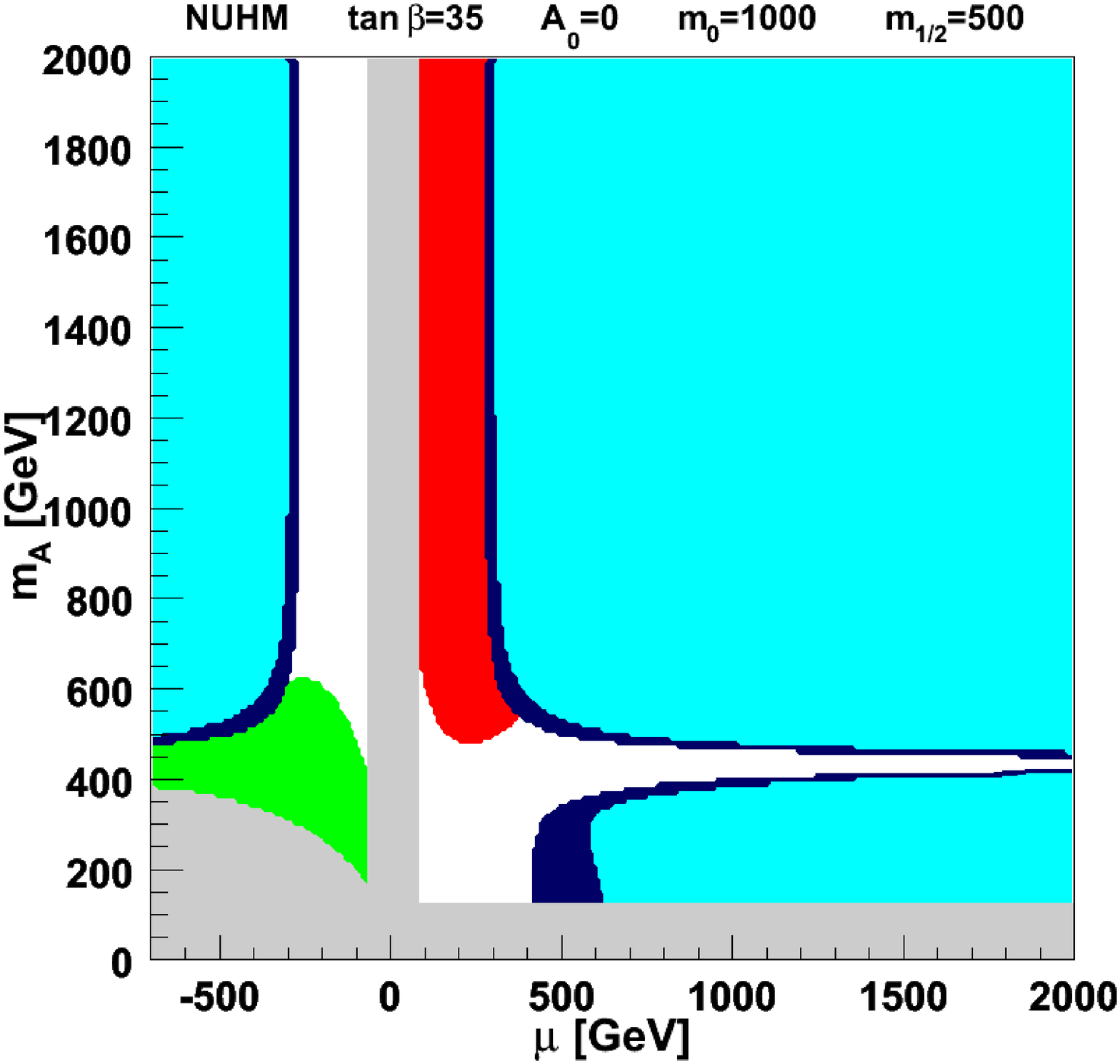}
\end{center}
\caption{Constraints on the mSUGRA (left) and NUHM (right) parameter planes. Contrary to the precedent figures, the light blue contours stand for the regions ``disfavored'' by the upper limit of the old dark matter constraints. The dark blue region together with the light blue, depicts the area disfavored by the upper limit of the new WMAP results. The other color conventions are as in the precedent figures. The only favored zone (by all the considered constraints) is therefore the white area.}\label{fig4}
\end{figure}%
In Fig. \ref{fig3}, we consider the NUHM parameter plane $(\mu, m_A)$, for $m_0=1000$ GeV, $m_{1/2}=500$ GeV, $\tan\beta=35$, $A_0=0$ and $\mu>0$ with the same color definitions as in the previous figure. First in this plane the LSP is not charged. The WMAP contours are small strips forming a cross. When increasing the expansion rate alterations, the cross becomes smaller and smaller, and in the last plot, for $n_D=8$ and $\kappa_D=10^{-5}$, the region favored by WMAP is extremely small, yet existing.\\
\\
The analysis presented here reveals that an increase of the expansion rate before BBN, even very small and presently undetectable in the cosmological observations, decreases the LSP freeze-out temperature, and increases the relic density by a factor of up to $10^6$. Even with a tiny increase, the WMAP favored regions in the SUSY parameter spaces are strongly displaced, so that they can become excluded by other constraints. Since the pre-BBN era is relatively unconstrained, and in many models beyond the cosmological standard model the expansion rate can be modified by the presence of a new fluid, a modified gravity, or any other reason, we consider that using the WMAP data to constrain SUSY should be done with caution. In particular, this study reveals that the relic density is increased when the expansion rate is modified by the presence of an extra density, but not decreased (or very slightly, in the case of a negative effective density). Thus, a relic density originally excluded by the lower WMAP limit could get increased and shifted to the permitted interval. Therefore, with such a modification of the cosmological expansion rate, only the upper limit of the WMAP constraints can be used safely, as it provides a limit on the maximum relic density. %
If we disregard the lower limit, we should not consider the region favored by WMAP anymore, but instead the region disfavored by the upper limit, as shown in Fig. \ref{fig4}. In this way, two problems are avoided: first, if it turns out that the cosmological standard model is too simple to describe correctly the pre-BBN era and that the expansion rate should be modified, our conservative upper limit (in the cosmological standard model) would still provide valid exclusion zones; second, if dark matter is composed of several components -- axions, dark fluids, ... -- and not only of the LSP, the upper limit of the WMAP constraint would still be reliable, contrary to the lower limit. Therefore, as a conclusion, we suggest using
\begin{equation}
\Omega_{\mbox{\small LSP}} h^2 < 0.135
\end{equation}
in order to explore SUSY parameter spaces with the WMAP data. Also, as we can see from Fig. \ref{fig4}, this WMAP constraint, together with the other usual constraints, can provide already good information on the favorite region of the SUSY parameter spaces.
\subsection*{Acknowledgments}
\noindent We would like to thank P. Salati, J. Rathsman, G. Ingelman and J. Edsj\"o for useful discussions and comments.

\end{document}